\documentclass[10pt,twocolumn]{article}
\usepackage[letterpaper,margin=0.75in,columnsep=0.25in]{geometry}
\usepackage{newtxtext,newtxmath}
\usepackage{graphicx}
\usepackage{floatrow}
\usepackage{float}
\usepackage{url}
\usepackage{orcidlink}
\usepackage{booktabs}
\usepackage{multirow}
\usepackage{xcolor,colortbl}
\usepackage{amsmath}
\usepackage[shortlabels]{enumitem}
\usepackage{hyperref}
\usepackage[super,comma,sort&compress]{natbib}
\hypersetup{
    colorlinks=true,
    linkcolor=blue,
    citecolor=blue,
    filecolor=blue,
    urlcolor=blue
}

\date{}

\makeatletter
\renewcommand{\fnum@figure}{\textbf{Figure \thefigure}}
\renewcommand{\fnum@table}{\textbf{Table \thetable}}
\makeatother

\newcommand{\tablecomments}[1]{\vspace{6pt}\noindent\footnotesize #1\normalsize}
\newcommand{\nwmax}{n_{W,\mathrm{max}}}
\newcommand{\ncmin}{n_{C,\mathrm{min}}}
\title{\bfseries \boldmath
The first three-dimensional map of thermal phases in the local interstellar medium
}
\author{
    Jonathan Shelest$^{1\ast}$\,\orcidlink{0009-0002-4021-3310},
    Shmuel Bialy$^{1}$\,\orcidlink{0000-0002-0404-003X},
    Troy A. Porter$^{2}$\,\orcidlink{0000-0002-2621-4440},
    Mark Wolfire$^{3}$\,\orcidlink{0000-0003-0030-9510},\\
    Matan Berko$^{1}$,
    Margarita Grinberg$^{4}$\,\orcidlink{0009-0008-4765-7452}
    \\
    \small$^{1}$Physics Department, Technion -- Israel Institute of Technology, Haifa 32000, Israel
    \\
    \small$^{2}$Kavli Institute for Particle Astrophysics and Cosmology, Stanford University, Stanford, CA 94305, USA
    \\
    \small$^{3}$Department of Astronomy, University of Maryland, College Park, MD 20742-2421, USA
    \\
    \small$^{4}$Faculty of Physics, Weizmann Institute of Science, Rehovot 7610001, Israel
    \\
    \small$^{\ast}$Corresponding author. Email: kyonatan@campus.technion.ac.il
}
\begin{document}
\maketitle
{\boldmath \bfseries
\noindent
The thermal state of interstellar gas controls whether it remains
warm and diffuse or cools into dense clouds. Twenty-one
centimetre observations have established a statistical picture of
neutral-gas phases, but their three-dimensional architecture has
not previously been mapped. Here we present
$\mathcal{P}_{\rm 3D}$, a three-dimensional reconstruction of
thermal phase in the local interstellar medium (ISM), covering a
1\,kpc-diameter region centred on the Sun and sampled on a
2\,pc Cartesian grid. 
$\mathcal{P}_{\rm 3D}$ combines high-resolution 3D maps of dust extinction and interstellar far-ultraviolet radiation with a neutral-ISM thermochemical model. The reconstruction reveals cold clouds
surrounded by thermally unstable envelopes and embedded in a
pervasive warm phase. Within $|z| \leq 150$\,pc of the Galactic
midplane, $\sim$41\% of the dust-traced neutral mass is thermally
unstable, implying phase cycling on a timescale of
$\sim$2.9--6\,Myr. Yet the cold phase
has a narrow density distribution, consistent with
internal Mach numbers that are transonic at most
($\mathcal{M}_s \lesssim 1.5$). Together, these findings
favour a dynamically cycling multiphase interstellar medium and motivate
reassessing star-formation models that assume a single-phase,
strongly supersonic cold-gas density field.
}

The interstellar medium (ISM) fills the space between stars with the gas
from which new stars and planetary systems form. Whether this gas
fragments into cold clouds or remains warm and diffuse is set by
its thermal state, the balance between heating by far-ultraviolet
(FUV) photons and cooling through line emission.
Classical theory predicts that the neutral ISM separates into two
thermally stable phases\cite{Field1969,Wolfire1995,Wolfire2003, Bialy2019}:
the cold neutral medium (CNM, $T\sim 200$ K), confined to dense clouds cooled by
[C\,{\sc ii}] and [O\,{\sc i}] emission, and the warm neutral
medium (WNM, $T\sim 7000$ K), filling the intercloud volume. Between these stable
branches lies the unstable neutral medium (UNM), expected in
classical theory to be transient and to carry little
mass\cite{Field1965}. Yet 21\,cm observations and
magnetohydrodynamical simulations consistently recover a
substantial population of gas in intermediate thermal
states\cite{Heiles2003,Murray2018,Audit2005,Saury2014,marchal2021},
pointing to an ISM continuously stirred by turbulence on
timescales of order of the thermal relaxation
time\cite{Saury2014,Gazol2013}.

\begin{figure*}
\centering
\includegraphics[width=1.0\textwidth]{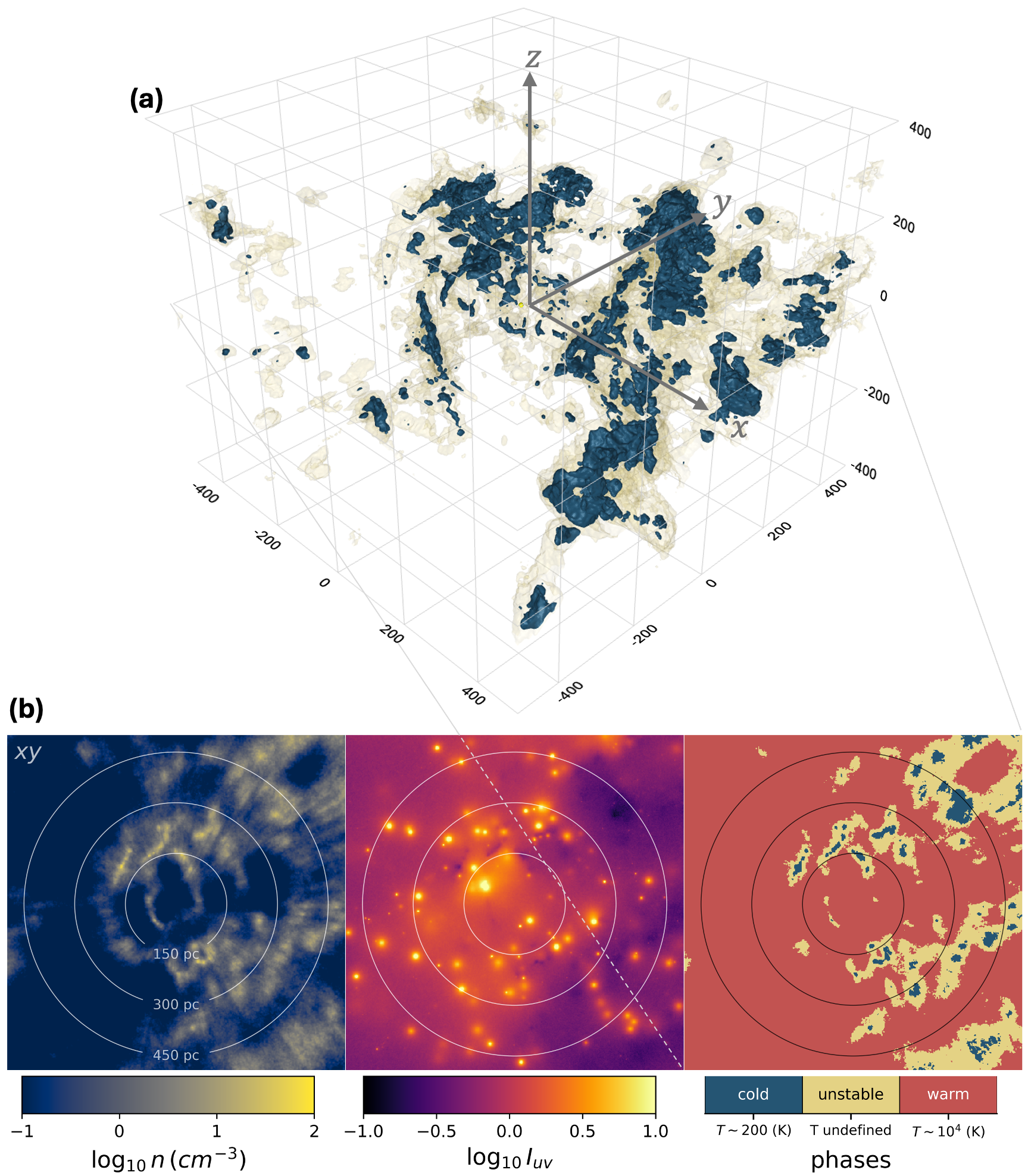}
\caption{\textbf{Three-dimensional thermal phase architecture of
the local ISM.}
\textbf{(a)} Three-dimensional rendering of the cold (blue) and
unstable (yellow) isosurfaces, with warm gas shown transparent.
The transparent warm volume traces WNM together with
unmodelled hot ionised gas, which carries negligible mass
(Methods). An interactive version is
available at
\href{https://sb2580.github.io/phases_3d/}{https://sb2580.github.io/phases\_3d/}.
\textbf{(b)} Slices through the Galactic plane ($z=0$) of gas
density $n$ (left), FUV radiation intensity $I_{\rm UV}$ (middle),
and phase classification (right). Concentric circles mark
heliocentric radii of 150, 300, and 450\,pc.}
\label{fig:slices_z0}
\end{figure*}

\begin{figure*}
\centering
\includegraphics[width=1.0\textwidth]{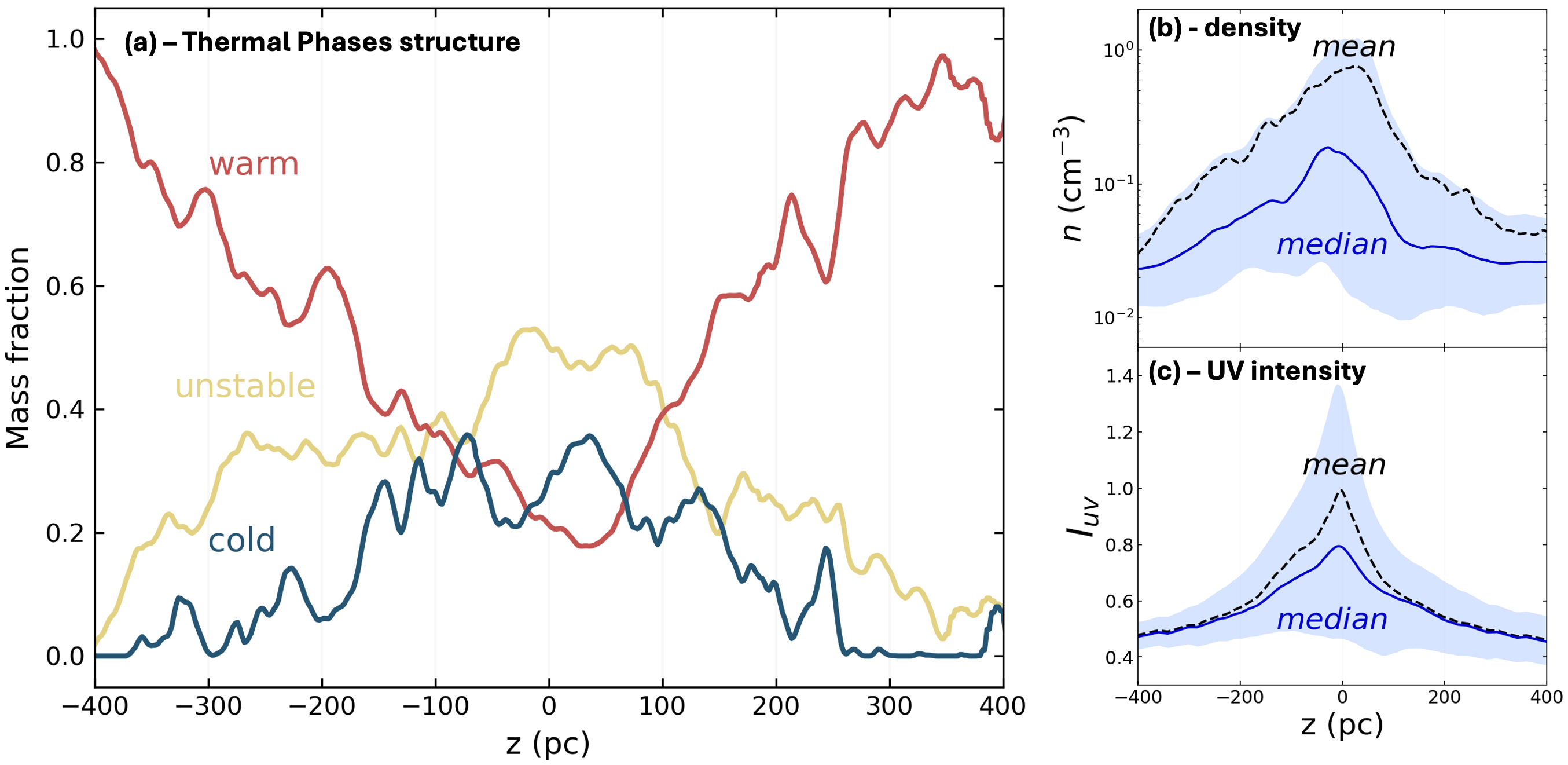}
\caption{\textbf{Vertical structure and thermal phase composition
of the local ISM.}
\textbf{(a)} Mass fractions of the warm, unstable, and cold
categories as functions of Galactic
height $z$. Averaged over $|z| \leq 150$\,pc, the three
categories carry comparable mass: $\langle f_C \rangle \approx
0.27$, $\langle f_U \rangle \approx 0.41$, $\langle f_W \rangle
\approx 0.32$. Warm gas becomes the largest mass
component for $|z| \gtrsim 150$\,pc. Fluctuations with
characteristic spacing $\Delta z \approx 50$--$100$\,pc reflect discrete
cloud complexes. \textbf{(b)} Mean (dashed) and median (solid)
gas density, with 15--85 percentile range (shaded), versus $z$.
\textbf{(c)} Same for the FUV radiation intensity $I_{\rm UV}$.}
\label{fig:phases_z}
\end{figure*}

Our observational knowledge of ISM thermal phases comes almost
entirely from 21\,cm hyperfine observations of atomic hydrogen in
emission and
absorption\cite{Heiles2003,Murray2018,Dickey1978,Dickey2002,Murray2020,marchal2019,marchal2021}.
These surveys provide powerful statistical constraints, but they
are intrinsically projected along the line of sight. They cannot
determine how cold, unstable, and warm gas are arranged in three
dimensions, how their abundances vary with Galactic height in the
solar neighbourhood, or how cold clouds connect geometrically to
their surroundings.
Over the past decade, Gaia parallaxes combined with photometric
surveys have driven a methodological shift from
sightline-integrated to volumetric mapping of the local ISM.
Parsec-scale three-dimensional dust extinction
maps\cite{Green2019,Lallement2019,Leike2020a,Edenhofer2024} now
resolve the spatial structure of nearby interstellar gas and have
enabled discoveries such as the Radcliffe Wave\cite{Alves2020},
the Per-Tau shell\cite{Bialy2021}, the Local Bubble as the
trigger of nearby star formation\cite{Zucker2022}, and the EoS
cloud\cite{Burkhart2025}. But dust traces where the gas is, not
which thermal state it occupies. The three-dimensional
distribution of thermal phase has therefore remained unmapped.

Here we present $\mathcal{P}_{\rm 3D}$, a volumetric
reconstruction of thermal phases in the local ISM.
$\mathcal{P}_{\rm 3D}$ synthesises three Gaia-era capabilities:
the Edenhofer et al.\cite{Edenhofer2024} three-dimensional dust
density map; a new three-dimensional FUV radiation field that
resolves individual OB stars within 1\,kpc; and a detailed
neutral-ISM thermochemical model\cite{Bialy2019}. Combining these
components, we assign a thermal-stability category to every cell
on a common 2\,pc grid spanning the solar neighbourhood. This
provides information that traditional sightline-integrated
diagnostics cannot access: the spatial architecture of thermal
phases, their vertical stratification, and the
volumetric density probability density function (PDF) of the cold phase. The last is
particularly interesting, since the cold-gas density PDF is a
central input to models of star formation in a turbulent ISM.

\section*{Results}

\subsection*{A three-dimensional thermal-stability map}
From the local gas density $n$ and FUV radiation intensity
$I_{\rm UV}$ (FUV intensity in \citep{Draine1978} units), $\mathcal{P}_{\rm 3D}$ assigns every cell in the
solar neighbourhood to a cold, warm, or unstable thermal state. The cold and warm states are the two stable,
near-isothermal branches of the thermochemical model, the
familiar CNM ($T\approx100$--$300$\,K) and WNM
($T\approx6{,}000$--$9{,}000$\,K). The unstable state is
the intermediate-density gas between them, collecting two
physically distinct populations: gas in heating-cooling balance
but on the thermally unstable branch (the classical UNM), and gas
out of balance entirely, in transition between the warm and cold
phases. We refer to cold and warm as phases and to the unstable
gas as a category, since it does not correspond to a single
stable branch. Because the density field is dust-based, the
cold phase may also include some H$_2$-associated gas, while
ionised gas, unmodelled here, fills volume but carries negligible
mass\cite{Ferriere2001,Draine2011} (Methods).

Fig.~\ref{fig:slices_z0}a shows the inferred cold and unstable
isosurfaces in three-dimensional space, and
Fig.~\ref{fig:slices_z0}b shows Galactic-plane slices of gas
density, FUV radiation field, and phase classification. The gas
density is highly inhomogeneous, with dense structures embedded
in diffuse material spanning orders of magnitude in density. The
FUV radiation field is also spatially structured, with peaks near
OB associations and local suppressions inside dust envelopes, but
its dynamic range is much narrower than that of the gas density.

The inferred phase map reveals a coherent multiphase architecture:
warm gas occupies most of the ambient volume, while cold clouds
are surrounded by unstable envelopes. Two effects set where cold
gas forms: high densities increase cooling and favour the cold
stable branch, while the dust columns toward these regions
attenuate the FUV field and suppress photoelectric heating. These coupled effects explain the spatial
coincidence of cold gas with regions of high density and low FUV
intensity in Fig.~\ref{fig:slices_z0}. The unstable gas is not scattered but organised into coherent
envelopes wrapping the cold structures, the spatial transition
from dense cold clouds to the diffuse warm surroundings.

\begin{figure*}
\centering
\includegraphics[width=\linewidth]{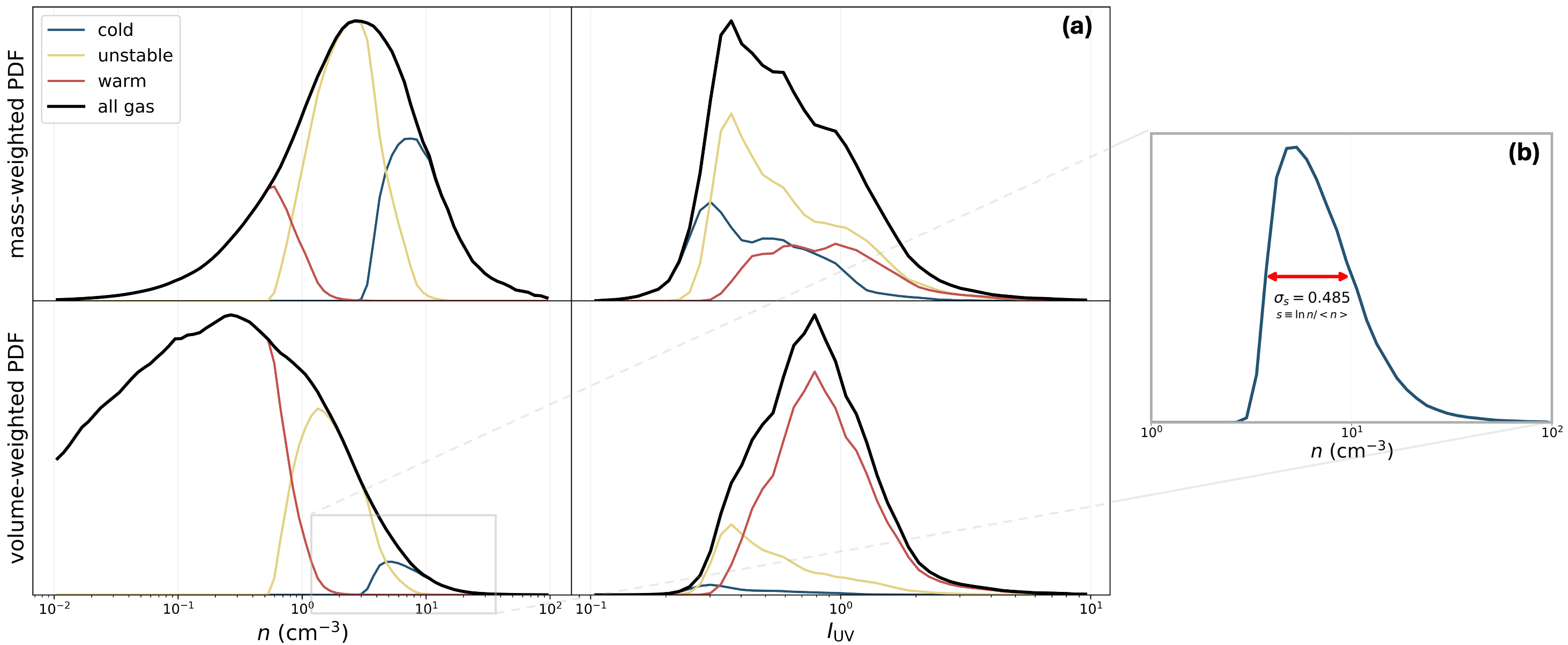}
\caption{\textbf{Phase-decomposed density and FUV intensity
PDFs.}
\textbf{(a)} Phase-decomposed PDFs of gas density $n$ (left
column) and FUV radiation intensity $I_{\rm UV}$ (right column),
mass-weighted (top row) and volume-weighted (bottom row), computed
within a midplane slab $|z| \leq 30$\,pc. In mass, the cold and
unstable categories dominate; in volume, the warm category
dominates. \textbf{(b)} Zoomed view of the volume-weighted
cold-phase density PDF. The logarithmic width is
$\sigma_s \equiv \sigma[\ln(n / \langle n \rangle)] = 0.485$,
implying an internal sonic Mach number
$\mathcal{M}_s = 0.5$--$1.5$ for the full range of turbulent
driving parameters (Eq.~\ref{eq:variance}).}
\label{fig:pdfs}
\end{figure*}

\subsection*{Vertical phase balance}

Fig.~\ref{fig:phases_z}a shows the mass fractions of the three
categories as a function of distance from the Galactic midplane.
Averaged over the central $|z| \leq 150$\,pc layer, the cold,
unstable, and warm mass fractions are
comparable: $\langle f_C \rangle \approx 0.27$,
$\langle f_U \rangle \approx 0.41$, and $\langle f_W \rangle
\approx 0.32$ (Table~\ref{tab:statistics_and_phases}). The
unstable gas is the largest single component from the midplane
out to $|z| \sim 100$\,pc. With increasing $|z|$, the gas becomes
progressively warmer, and the warm gas dominates for
$|z| \gtrsim 150$\,pc.
This trend follows from how the gas density and FUV field vary
with height (Fig.~\ref{fig:phases_z}b,c; Extended Data
Fig.~\ref{fig:iuv_n_pdf}). The density falls steeply away from the
midplane, whereas the FUV field declines far more gradually,
because it averages over many luminous sources spread across the
disk while the density traces local material. As cooling weakens
relative to heating, the phase balance shifts toward the warm
phase.

Superposed on this smooth decline are fluctuations in the cold
and unstable fractions on scales of $\Delta z \sim 50$--$100$\,pc,
tracing individual cold-cloud complexes. The same scale appears
in the spacing of FUV peaks around local sources
(Fig.~\ref{fig:slices_z0}b) in agreement with theoretical models\cite{Bialy2020}. The vertical phase
structure thus reflects both the global density decline and the
discrete clouds and sources of the solar neighbourhood.

\subsection*{A substantial unstable reservoir}
The vertical decline in density explains why the warm phase
becomes increasingly dominant away from the midplane, but it does
not by itself explain why the midplane contains so much unstable
gas. Across the central $|z| \leq 150$\,pc layer, $\langle f_U
\rangle \approx 0.41$ of the dust-traced neutral mass lies in the
unstable category, with a systematic spread of $\approx 0.05$
across thermochemical-model and dust-map choices (Methods). This contrasts with the classical two-phase
picture in which thermal instability segregates gas onto the
cold and warm stable branches, leaving the intermediate regime
nearly empty\cite{Field1969,Field1965}.

This unstable reservoir implies a medium stirred on a timescale of order the thermal relaxation time. In a simple steady-state cycling balance (Methods), the ratio $f_U/(1-f_U)$ sets the rate of dynamical injection into the unstable regime relative to the thermal relaxation rate out of it; for $\langle f_U \rangle \approx 0.41$ and $t_{\rm cool} \approx 2$\,Myr\cite{Hennebelle1999,Wolfire2003,Bialy2019} this gives a cycling time $t_{\rm dyn} \approx 2.9$\,Myr. The local ISM therefore lies in the $t_{\rm dyn} \sim t_{\rm cool}$ regime in which turbulence and large-scale flows maintain large, long-lived unstable populations\cite{sanchez-salcedo2002,Audit2005,Saury2014,Gazol2013,Gazol2016}. Limited spatial resolution can smooth small cold structures into the intermediate-density interval, biasing $f_U$ high (Methods); a lower true value would lengthen the cycling time, but only to $t_{\rm dyn} \approx 6$\,Myr, still of order $t_{\rm cool}$, so the actively cycling picture holds.

\subsection*{The cold-phase density PDF}

Beyond the mass fractions, the per-category density and FUV PDFs characterise the internal structure of each phase.
Fig.~\ref{fig:pdfs}a shows the mass- and volume-weighted PDFs of
gas density and FUV intensity for the three categories, computed
within a midplane slab $|z| \leq 30$\,pc. 
Cold and unstable gas dominate the mass while warm gas dominates the
volume.
Overall, the density PDF spans several
orders of magnitude, whereas $I_{\rm UV}$ is comparatively
narrow, reflecting the spatial smoothing of FUV radiation from
many stellar sources across the solar neighbourhood.

The density PDF of the cold phase is particularly important, since
this phase is the one most susceptible to gravitational collapse.
Canonical isothermal-turbulence models of star formation predict
the star-formation efficiency from this
PDF\cite{Krumholz2005,McKee2007,padoan2011,Federrath2012,burkhart2018},
relating its width to the internal sonic Mach number through the
density-variance relation\cite{Padoan1997,Federrath2010}
\begin{equation}
\sigma_s^2 = \ln\!\left(1 + b^2 \mathcal{M}_s^2\right),
\label{eq:variance}
\end{equation}
where $\sigma_s$ is the standard deviation of $s \equiv \ln(n /
\langle n \rangle)$, and $b \in [1/3,1]$ parametrises the
turbulent driving mode\cite{Federrath2010}  (Methods).

$\mathcal{P}_{\rm 3D}$ enables a direct measurement of the
cold-phase density PDF (Fig.~\ref{fig:pdfs}b). We find
$\sigma_s = 0.485$, which inverts to $b\mathcal{M}_s = 0.5$. For
fiducial mixed driving $b = 0.4$\cite{Federrath2010} this gives
$\mathcal{M}_s \approx 1.3$, and across the full range of driving
the Mach number spans $\mathcal{M}_s = 0.5$ $(b=1$, compressive$)$
to $1.5$ $(b=1/3$, solenoidal$)$. On the resolved scales of the reconstruction, the cold phase is therefore internally sub- to transonic for all allowed driving modes (Methods).

\section*{Discussion}

$\mathcal{P}_{\rm 3D}$ adds a new, spatially resolved view of the
local neutral ISM, complementing line-of-sight phase studies with
a three-dimensional multiphase architecture built from Gaia-based
3D maps. The medium is organised into cold clouds enveloped by unstable
transition layers, with warm gas filling the ambient volume.
Two quantitative results stand out: a large unstable mass
reservoir near the midplane, and a narrow density distribution
within the cold phase. We assess both below, first against the
21\,cm phase studies and then as a constraint on the turbulent
state of the cold gas.

\subsection*{Comparison with 21\,cm observations}
The multiphase composition of the neutral ISM has long been
established by 21\,cm emission-absorption
surveys\cite{Heiles2003,Murray2018,McClure-Griffiths2023}, but a
direct comparison with $\mathcal{P}_{\rm 3D}$ is not
straightforward, for three reasons. First, the classifications
differ. 21\,cm surveys separate phases by temperature, whereas we
separate by thermal stability, so our unstable category contains
both the classical unstable gas they identify and gas actively
transitioning between phases, which has no single temperature.
Second, 21\,cm absorption constrains the kinetic temperature well
only for cold gas, where the line is strong and the spin
temperature tracks the kinetic temperature; at higher temperatures
both conditions weaken, so unstable and warm gas are hard to
separate. Third, 21\,cm sightlines integrate over the full
vertical column, where the cold gas occupies a smaller scaleheight
than the warm gas\cite{Kalberla2009,soler2022}, so projected
fractions are warm-weighted relative to our local
$|z| \leq 150$\,pc slab.

These limitations all act on the warm and unstable gas, leaving
the cold fraction as the cleanest comparison. Our midplane value
$\langle f_{\rm CNM} \rangle \approx 0.27$ ($|z| \leq 150$\,pc) is
at the lower end of the reported observational range
$f_{\rm CNM}\approx0.28$--$0.40$
\cite{Heiles2003,Murray2018,McClure-Griffiths2023}.
Because $\mathcal{P}_{\rm 3D}$ is dust-based, its cold gas may
include some H$_2$-associated material absent from 21\,cm, a minor
effect for the diffuse, mostly atomic local ISM (Methods).

Our unstable fraction is higher than 21\,cm-based UNM values,
$\langle f_U \rangle \approx 0.41$ against $0.20$--$0.28$\cite{Murray2018,McClure-Griffiths2023}.
This is expected: our category is broader, counting transitioning
gas the temperature-based UNM omits, and 21\,cm is least reliable
in exactly this regime (see above). Our value may in
addition be inflated if limited resolution smooths cold
substructure into the intermediate-density interval (Methods). Even adopting a floor as low as the 21\,cm estimate, $f_U \approx 0.24$, the implied cycling time rises only to $t_{\rm dyn} \approx 6$\,Myr, still of order the thermal relaxation time.

\subsection*{Internal turbulent state of the cold phase}
Single-phase isothermal star-formation models assume the cold
star-forming reservoir is strongly supersonic, with a broad
lognormal density
PDF\cite{Krumholz2005,McKee2007,padoan2011,Federrath2012,burkhart2018}.
In contrast, the cold-phase PDF in
$\mathcal{P}_{\rm 3D}$ is narrow, corresponding to internal
sub- to transonic Mach numbers. 
This
agrees with multiphase turbulence simulations in which cold
structures are internally quiescent while moving at larger bulk
velocities through the warm and unstable
medium\cite{Audit2005,Saury2014,Kobayashi2022}. The higher Mach
numbers from non-thermal CNM linewidths,
$\mathcal{M}_s \approx 3$\cite{Heiles2003}, may therefore reflect
bulk motion between cold structures superposed along the sightline
rather than turbulence within a single one.
The same
large-scale motions that keep a substantial mass thermally
unstable would then appear as bulk velocities between internally
quiescent cold structures.

\subsection*{The emerging picture}
Together, the large unstable reservoir and the narrow
cold-phase density PDF favour a dynamically cycling multiphase
ISM, in which the cold star-forming reservoir is not an isolated
turbulent medium but one component of a larger multiphase flow.
This challenges the direct application of single-phase, strongly
supersonic isothermal density PDFs to the cold neutral gas from
which molecular clouds assemble. $\mathcal{P}_{\rm 3D}$ therefore
provides a volumetric observational benchmark for multiphase
simulations and for star-formation models that seek to connect
diffuse atomic gas to molecular-cloud formation.

\clearpage
\bibliographystyle{naturemag}
\bibliography{references}

\section*{Acknowledgments}
We thank Benjamin Godard for a thorough reading of the manuscript
and for detailed comments that significantly improved the paper.
S.B.\ acknowledges support from the ISF grant number 2071540, the
GIF grant number I-1568-303.7/2024, the NSF-BSF grant number
2023761, and the Alon Fellowship prize for junior faculty.

\section*{Author contributions}
J.S.\ led the project, constructed $\mathcal{P}_{\rm 3D}$,
performed the thermal phase analysis and statistical
characterisation, and produced all figures except
Extended Data Figs.~\ref{fig:phase_classification} and
\ref{fig:iuv_n_pdf}.
S.B.\ conceived and supervised the project, provided the BS19 thermochemical model, and contributed to the scientific interpretation. T.A.P.\ developed and provided the
three-dimensional FUV radiation field model and contributed to
the ISRF methodology. M.W.\ provided the Wolfire-type
thermochemical models used in the robustness assessment. M.B.\
contributed to the data analysis and produced
Extended Data Figs.~\ref{fig:phase_classification} and
\ref{fig:iuv_n_pdf}. M.G.\ contributed to the thermochemical
modelling and thermal phase analysis. S.B.\ and J.S.\ wrote the
manuscript. All authors provided scientific comments and approved
the final manuscript.
\section*{Competing interests}
The authors declare no competing interests.

\section*{Data availability}
The $\mathcal{P}_{\rm 3D}$ data product will be publicly
available as a zarr datacube on Zenodo. The three-dimensional dust map is publicly
available at \href{https://zenodo.org/records/10658339}{https://zenodo.org/records/10658339}. Summary statistics and phase
mass fractions at selected heights are provided in
Table~\ref{tab:statistics_and_phases}.

\section*{Code availability}
Code used to construct $\mathcal{P}_{\rm 3D}$, compute the
thermal phase classification, generate all figures, and perform
the robustness tests, will be deposited on Github upon publication.
The thermochemical model of Bialy and Sternberg\cite{Bialy2019}
is available from S.B.\ upon reasonable request.

\setcounter{figure}{0}
\setcounter{table}{0}

\section*{Methods}

\textbf{Three-dimensional dust map.}
We use the three-dimensional dust map of Edenhofer et
al.\cite{Edenhofer2024} (E24), which provides differential
extinction in three dimensions at parsec-scale resolution within
1.25\,kpc of the Sun. The map is built from stellar distance and
extinction estimates of Zhang et al.\cite{Zhang2023}, whose
data-driven approach combines Gaia BP/RP spectra, near-infrared
photometry from 2MASS\cite{Skrutskie2006}, unWISE
photometry\cite{Wright2010a,Schlafly2019a}, and spectroscopic
data from LAMOST\cite{Wang2022,xiang2022}, enabling simultaneous
forward-modelling of stellar extinction, distance, and intrinsic
parameters ($T_{\rm eff}$, [Fe/H], $\log g$).
We convert the differential extinction $A'_{\rm ZGR23}$
(mag\,pc$^{-1}$) to gas volume density by interpolating the
Zhang et al.\ extinction curve at a reference wavelength of
$0.673\,\mu$m and applying the Draine et al.\cite{Draine2007}
extinction cross-section ($\sigma_{\rm ext} = 3.52 \times
10^{-22}$\,cm$^2$) together with a scaling factor of 2.04:
\begin{equation}
n_{\rm H} = 1727\, A'_{\rm ZGR23}\;\text{cm}^{-3}.
\label{eq:dust_to_gas}
\end{equation}
This calibration agrees to within $\sim$5\% with the independent
dust-to-gas conversion of O'Neill et al.\cite{ONeill2024}.

\textbf{Three-dimensional FUV radiation field.}
The 6--13.6\,eV FUV component of the interstellar radiation field
is the primary driver of photoelectric heating and therefore of
the neutral ISM temperature. State-of-the-art three-dimensional
ISRF models\cite{porter2017} (F98, R12) assume smooth stellar and
dust distributions, which limits their UV predictions in two
ways: they lack small-scale structure, and they do not resolve
individual hot, luminous stars that dominate UV output. 

We
therefore construct a new three-dimensional FUV field for the
solar neighbourhood, following the methodology developed in
parallel by T.~Porter et al.\ (in prep.; hereafter P25), based on
\cite{PorterICRC}, and applied here at full radiative-transfer
detail to our analysis volume.
O and B stars within 1\,kpc are treated as individual point
sources, using the Hipparcos catalogue\cite{vanleeuwen2007} and
the Gaia golden OBA sample\cite{gaiacollaboration2023}. 
To avoid
double-counting stars present in both catalogues while preserving
the optimal astrometric solution for each distance regime,
samples are matched based on count histograms binned by
heliocentric distance. Hipparcos sources with complete MK
spectral type and luminosity class information are assigned
spectral energy distributions from the groupings in Robitaille et
al.\cite{robitaille2012}; for Gaia sources, photospheric
parameters ($T_{\rm eff}$, $\log g$) are transformed to MK
spectral types using calibration tables from Straizys and
Kuriliene\cite{straizys1981} and Cox\cite{cox2000}. Stars with
ambiguous or missing classifications default to main sequence
with the least luminous available type, conservatively avoiding
overestimation of UV flux. O stars are assigned to the broad
``Young OB'' spectral model.

Radiative transfer is performed on the high-resolution E24 dust
map within 1\,kpc, transitioning smoothly to the F98 large-scale
dust model at larger distances to avoid discontinuities. At
wavelengths beyond the FUV, we retain the F98 background, as F98
and R12 converge closely at optical and far-infrared wavelengths.
The resulting FUV field captures realistic spatial variations on
scales of $\sim$10--100\,pc, driven by individual luminous stars
and inhomogeneous small-scale dust.


\textbf{Thermochemical model.}
Given local values of gas density $n$ and FUV intensity
$I_{\rm UV} \equiv F_{\rm UV} / F_{\rm UV,\odot}$, where
$F_{\rm UV,\odot} = 2.7 \times
10^{-3}$\,erg\,cm$^{-2}$\,s$^{-1}$\cite{Draine1978,Bialy2021},
we compute the steady-state gas temperature using the
thermochemical model of Bialy and Sternberg\cite{Bialy2019}
(BS19). The model solves simultaneously for the abundances of key
species (H, H$_2$, e$^-$, C, C$^+$, CO) and the gas temperature
by requiring chemical and thermal steady state across the neutral
ISM ($T \approx 10$--$10^4$\,K). 
Heating processes include
photoelectric heating by FUV radiation, cosmic-ray ionisation
heating, and chemical reaction heating from H$_2$ formation,
photodissociation, and photo-pumping. Cooling processes include
H\,Ly$\alpha$, [C\,{\sc ii}]\,158\,$\mu$m, [C\,{\sc i}],
[O\,{\sc i}]\,63\,$\mu$m, H$_2$ rovibrational emission,
dust-assisted recombination cooling, and thermal dust emission.

The model takes four parameters: $n$, $I_{\rm UV}$, metallicity
$Z'$, and cosmic-ray ionisation rate $\zeta$. We adopt fiducial
values $Z' = 1$ and $\zeta = 10^{-16}$\,s$^{-1}$. The
steady-state temperature is weakly sensitive to both: $Z'$ enters
similarly into photoelectric heating and fine-structure line
cooling, leaving their balance, and hence the temperature,
nearly unchanged, while $\zeta$ enters mainly through its effect
on the electron fraction at low densities (see BS19 for
details). With $Z'$ and $\zeta$ fixed, the steady-state   pressure
curve is determined by $n$ and $I_{\rm UV}$.

\textbf{Phase classification.}
For each $I_{\rm UV}$, the steady-state pressure curve
$P_{\rm eq}/k_{\rm B}=nT_{\rm eq}(n)$ is S-shaped, with a local
maximum at $\nwmax(I_{\rm UV})$ and a local minimum at
$\ncmin(I_{\rm UV})$ bracketing the thermally unstable branch
where $dP/dn<0$ (Extended Data  Fig.~\ref{fig:phase_classification}). Both turning points exist across the range of
$I_{\rm UV}$ relevant for our analysis volume.
Over $0.1 \le I_{\rm UV}\le 10$, the two turning points are well
approximated by power laws,
\begin{align}
\nwmax(I_{\rm UV}) &\approx I_{\rm UV}^{0.51}\,{\rm cm}^{-3}, \\
\ncmin(I_{\rm UV}) &\approx 6.61\,I_{\rm UV}^{0.53}\,{\rm cm}^{-3},
\end{align}
which provide a convenient closed-form approximation to the phase
boundaries. For the analysis presented here we do not use these fits
but instead solve $dP/dn=0$ exactly at each $I_{\rm UV}$; the resulting
separators are the curves shown in the bottom panel of Extended Data
Fig.~\ref{fig:phase_classification}, and the two turning points marked
in the top panel are the corresponding values at $I_{\rm UV}=1$. We
classify each volume cell into one of three categories:
\begin{align}
\text{Warm:}     &\quad n < \nwmax(I_{\rm UV}), \\
\text{Unstable:} &\quad \nwmax(I_{\rm UV}) < n < \ncmin(I_{\rm UV}), \\
\text{Cold:}     &\quad n > \ncmin(I_{\rm UV}) .
\end{align}

This definition is geometric in the $(n,I_{\rm UV})$ plane and
does not require an observed kinetic temperature. Temperature
enters only as an output of the thermochemical model. For cells
in the warm or cold categories, the gas is assigned to one of the two
stable equilibrium branches, and the corresponding 
temperature is physically meaningful within the steady-state
model. These stable branches are near-isothermal over the
radiation-field range relevant to the local volume, with
$T_{\rm eq}\approx6{,}000$--$9{,}000$\,K on the warm branch (WNM) and
$T_{\rm eq}\approx100$--$300$\,K on the atomic cold
branch (CNM). Molecular gas, when present in the cold phase, can be
colder, but it is not separated as a distinct category in this
work.

The unstable category includes two types of gas, (1) the
classical thermally unstable equilibrium branch where local
heating and cooling balance but the solution is unstable, yielding intermediate temperatures, $T\approx300$--$6000$\,K, and (2) gas displaced from equilibrium entirely with heating and cooling rates that are unbalanced. Such off-equilibrium gas is expected in shock-compressed and turbulent multiphase flows\cite{Hennebelle1999,Audit2005,Godard2024}. 
As such, the unstable category is not a formal phase of the gas, and does not have a unique kinetic temperature.

The existence of type-(2) gas raises the converse question: some gas counted as warm or cold may itself be off-equilibrium, sitting at a stable density but not on the $P(n)$ curve, so $f_U$ is formally a lower limit on the off-equilibrium mass. We expect this missed mass to be small. Above $\ncmin$ the cooling time is short, of order 10 kyr\cite{Bialy2019}, far shorter than the dynamical time, so gas displaced to high density relaxes onto the cold branch and is correctly counted as cold. In the warm regime, shock-driven compression in converging flows displaces gas into the unstable interval, where it is counted, rather than leaving it off-equilibrium at warm densities. The dominant systematic on $f_U$ acts in the opposite direction: finite spatial resolution, assessed below (Resolution and the unstable mass fraction).

\textbf{Molecular and ionised gas.}
The density field used here is dust-based and therefore traces
gas associated with dust, not H\,{\sc i} alone. The cold phase
corresponds to the stable CNM branch of the neutral
thermochemical model, but it can also include material associated
with H$_2$. Within BS19, H$_2$ chemistry is solved in steady
state and contributes to the heating-cooling balance, but
molecular gas is not separated as a distinct category. This
distinction is important when comparing to 21\,cm surveys, which
measure atomic gas. We therefore do not interpret the comparison
with 21\,cm CNM fractions as a one-to-one measurement of the same
quantity.

We expect the molecular contribution to have a limited effect on the comparison with 21\,cm (Discussion). The diffuse local
neutral ISM is dominated by atomic gas, while H$_2$ is
concentrated in dense molecular interiors. In addition,
stellar-extinction-based 3D dust maps under-recover the most
opaque molecular cores because background stars become sparse
behind high-extinction regions. Thus the mass traced by
$\mathcal{P}_{\rm 3D}$ is weighted toward the atomic and
translucent envelopes of nearby clouds, not toward a complete
census of dense molecular interiors. Future work combining
$\mathcal{P}_{\rm 3D}$ with CO, dust-emission, and H$_2$ tracers
can separate the molecular component explicitly.

The warm phase corresponds to the stable WNM branch of the
neutral thermochemical model. We do not model ionised gas
self-consistently. The diffuse hot ionised medium has a density
far below $\nwmax$ and is therefore absorbed into the warm
category, where it occupies a large volume fraction but
contributes negligible mass\cite{Ferriere2001,Draine2011}. Denser
ionised gas, such as discrete H\,{\sc ii} regions, can in
principle reach the intermediate or high densities of the
unstable and cold categories, but its mass and volume within the
analysis region are negligible. The warm phase should therefore
be interpreted as the stable WNM solution for neutral gas, not as
a complete census of all gas occupying the same volume.


\textbf{Constructing $\mathcal{P}_{\rm 3D}$.}
We interpolate both $n_{\rm H}$ and $I_{\rm UV}$ onto a common
Cartesian heliocentric grid with 2\,pc cells, spanning
$x,y \in [-0.5,0.5]$\,kpc and $z \in [-0.4,0.4]$\,kpc, with the
Sun at the origin. The 2\,pc spacing is the voxel spacing of this
common grid; the effective spatial resolution is set by the
underlying dust reconstruction and is therefore scale- and
distance-dependent.
At each grid cell, we evaluate the BS19 thermochemical model for
the local $(n_{\rm H},I_{\rm UV})$ pair by interpolating within a
precomputed grid of equilibrium solutions and phase boundaries.
Each cell is thereby assigned to a category, yielding
$\mathcal{P}_{\rm 3D}$.
Unless otherwise stated, all reported phase fractions, PDFs, and
related statistics are computed over cells within a heliocentric
distance of 500\,pc, where the physical resolution of the dust
map is least affected by angular smoothing and line-of-sight
averaging; cells in the corners of the rectangular grid that lie
beyond this distance are excluded. Slab-averaged mass fractions
are computed by summing the mass in each category over all cells
with $|z| \leq z_{\max}$ and dividing by the total mass in the
slab.


\textbf{Dynamical cycling timescale.}
The mass-balance argument used in the Results is a two-compartment
rate equation for gas cycling between the unstable category and
the stable phases. In steady state,
\begin{equation}
\frac{f_U}{1 - f_U} \simeq \frac{k_{\rm dyn}}{k_{\rm cool}}
= \frac{t_{\rm cool}}{t_{\rm dyn}},
\label{eq:fU_balance}
\end{equation}
where $k_{\rm cool}=1/t_{\rm cool}$ is the thermal relaxation
rate and $k_{\rm dyn}=1/t_{\rm dyn}$ is the rate at which
dynamical motions cycle gas into the unstable regime. This has
the same physical content as the $\eta \equiv t_{\rm cool} /
t_{\rm dyn}$ parameterisation of Sanchez-Salcedo et
al.\cite{sanchez-salcedo2002}, who showed that order-unity
$\eta$ sustains large, long-lived populations of unstable gas.
For $\langle f_U \rangle \approx 0.41$ in the central
$|z| \leq 150$\,pc slab, Eq.~(\ref{eq:fU_balance}) gives
$t_{\rm cool}/t_{\rm dyn} \approx 0.7$, or
$t_{\rm dyn} \approx 1.4\,t_{\rm cool}$.
The adopted cooling time $t_{\rm cool} \approx 2$\,Myr is robust
across independent estimates. Wolfire et al.'s\cite{Wolfire2003}
standard model for solar-neighbourhood conditions ($N_c \approx
10^{20}$\,cm$^{-2}$, $G_0 \approx 1.7$) gives
$t_{\rm cool} \approx 2.3$\,Myr. 

The cooling length
$\lambda_{\rm cool} \approx 18$\,pc of Hennebelle and
Perault\cite{Hennebelle1999} divided by the warm-phase
sound speed gives $\approx 2.2$\,Myr. The reverse process,
heating from the unstable regime back toward the warm phase,
gives a consistent timescale: for an average photoelectric
heating rate per particle $\Gamma_{\rm PE} = 2 \times
10^{-26}$\,erg\,s$^{-1}$ and a warm-phase temperature
$T_W = 6{,}000$\,K, $t_{\rm heat} \approx (5/2) k_{\rm B} T_W /
\Gamma_{\rm PE} \approx 3.3$\,Myr, consistent with the numerical
results of Bialy and Sternberg\cite{Bialy2019}. The various
thermal relaxation estimates are consistent to within a factor of
$\approx$1.7.

Because $t_{\rm dyn} = t_{\rm cool}\,(1-f_U)/f_U$, a lower
unstable fraction, from potential smoothing of cold gas into the intermediate interval (Resolution and the unstable mass fraction), lengthens the cycling time: for $f_U$ from our measured $0.41$ down to $0.24$ (the central 21\,cm UNM estimates; see Discussion), $t_{\rm dyn}$ increases from $\approx 2.9$ to $\approx 6$\,Myr, remaining of order $t_{\rm cool}$.

\textbf{Volumetric density PDF and Mach number.}
To measure the volume-weighted cold-phase density PDF, we select
all $\mathcal{P}_{\rm 3D}$ grid cells classified as cold in a
midplane slab $|z| \leq 30$\,pc. We compute the
standard deviation of $s \equiv \ln(n/\langle n \rangle)$ and
obtain $\sigma_s = 0.485$. Inverting Eq.~(\ref{eq:variance})
yields $b\mathcal{M}_s$; the values for fiducial driving
$b=0.4$ and across the full physical range of $b$ are given in
the Results. The boundary of the cold phase in $(n,I_{\rm UV})$
space is set by the $\ncmin$ curve rather than by a hard density
cut, producing a soft left edge of the PDF that reflects the
physical phase definition.
Although Eq.~(\ref{eq:variance}) was derived for isothermal
turbulence, its application to the cold-phase density PDF is
supported by non-isothermal multiphase simulations. In
converging-flow simulations that resolve thermal instability and
the formation of a multiphase medium, Kobayashi et
al.\cite{Kobayashi2022,Kobayashi2023} found that the width of the
cold-only density PDF is reproduced by Eq.~(\ref{eq:variance})
when $\mathcal{M}_s$ is the velocity dispersion internal to cold
structures, rather than the bulk velocity between them. Because
our volume-weighted PDF is measured on the cold-phase density
field, the inverted $\mathcal{M}_s$ corresponds to this resolved
internal Mach number. A residual $\sim$10--20\% uncertainty in
the relation in the transonic regime\cite{Federrath2010} does
not affect the conclusion that the cold phase is
internally sub- to transonic.

Unresolved sub-parsec structure could broaden the true small-scale
cold-phase density PDF relative to the reconstructed one. Spatial
averaging and dust-map regularisation suppress sharp density
contrasts, dilute high-density peaks, and reduce the variance of
the recovered density field. The inferred Mach number therefore
characterizes the resolved cold-phase density field. It should
not be interpreted as excluding additional substructure below the
effective resolution of the 3D dust reconstruction.


\textbf{Resolution and the unstable mass fraction.}
The main systematic on $f_U$ is finite spatial resolution: cold
structures smaller than the effective resolution, if averaged with
lower-density gas within a voxel, can be counted in the
intermediate-density interval, biasing $f_U$ high. This acts
opposite to the phase-space lower limit discussed above, where
off-equilibrium gas at stable densities is missed. Three lines of
evidence bound the effect. First, the recovered $f_U(z)$ profile
is stable for analysis-volume cuts from 500 out to 600--800\,pc,
despite the physical resolution degrading with distance, and
departs from it only beyond $\sim$1\,kpc where line-of-sight
smoothing becomes severe. Second, the unstable gas forms coherent
envelopes around cold structures (Fig.~\ref{fig:slices_z0}),
consistent with multiphase MHD
simulations\cite{Bellomi2020,Saury2014}; beam averaging would
redistribute mass at cold-unstable boundaries, not produce these
large coherent envelopes. Third, the same simulations find most
cold mass in complexes larger than several
parsecs\cite{Bellomi2020,Saury2014}, so the bulk of the CNM is
well resolved here. Unresolved sub-parsec substructure would
therefore refine $f_U$ and $f_C$ without changing our conclusions;
in particular, the cycling time stays of order the thermal
relaxation time even at the most conservative $f_U$ (see Dynamical
cycling timescale, and the Discussion section). Future sub-parsec 3D dust maps will quantify
the cold mass on these scales.

\textbf{Other caveats and systematic uncertainties.}
Phase fractions vary with Galactic
environment\cite{Dickey2009,Kalberla2018,marchal2021}; the
values reported here are representative of the solar
neighbourhood and should not be extrapolated to spiral arms, the
inner Galaxy, or regions of active stellar feedback without
further analysis.
The phase classification is defined physically by the thermal
stability boundary $dP/dn = 0$, which marks where the stable warm
and cold solutions cease to exist; the equilibrium temperature is
an output of the thermochemical model, not an input to the
classification. We assess the sensitivity of the resulting mass
fractions to two sources of systematic uncertainty, the
thermochemical model and the three-dimensional dust map. For the
thermochemical model, we repeat the full analysis using
Wolfire-type models\cite{Wolfire2003} with
$\zeta = 1.6 \times 10^{-16}$, $8 \times 10^{-17}$, and
$4 \times 10^{-17}$\,s$^{-1}$. For the dust map, we apply the
BS19 model to the Leike et al.\cite{Leike2020a} map with the FUV
field recomputed self-consistently. All configurations reproduce
the same vertical structure (Extended Data
Fig.~\ref{fig:alt_models}); quantitative differences within the
central $|z| \leq 150$\,pc layer are $\delta f_C \approx 0.04$,
$\delta f_U \approx 0.05$, and $\delta f_W \approx 0.07$.

\setcounter{figure}{0}
\setcounter{table}{0}
\makeatletter
\renewcommand{\fnum@figure}{\textbf{Extended Data Fig. \arabic{figure}}}
\renewcommand{\fnum@table}{\textbf{Extended Data Table \arabic{table}}}
\makeatother
\onecolumn
\begin{center}
{\large\bfseries Extended Data}
\end{center}
\vspace{1em}
\begin{figure}[H]
\centering
\includegraphics[width=0.7\linewidth]{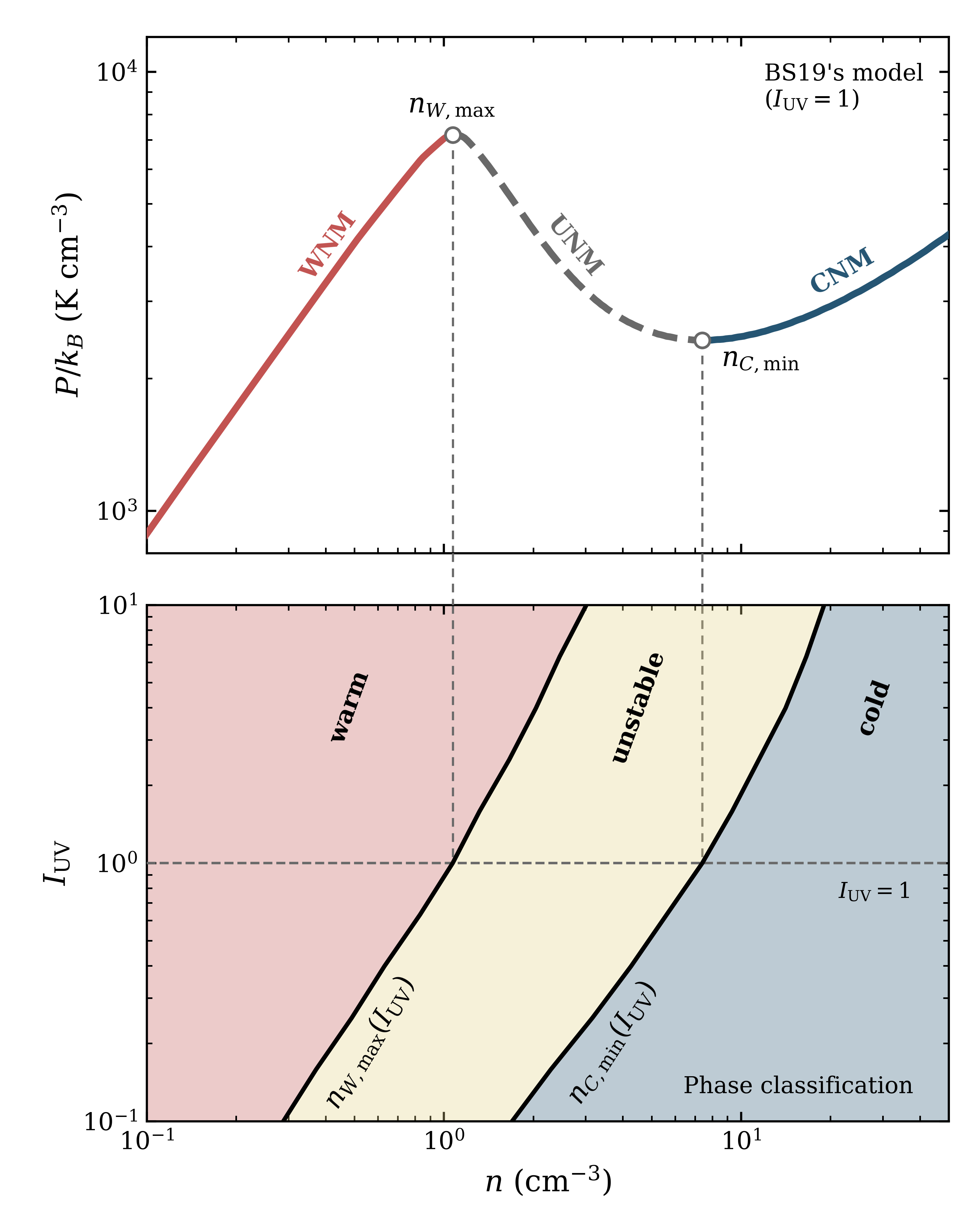}
\caption{\textbf{Thermochemical model and the $(n,I_{\rm UV})$
classification.}
\textit{Top:} Steady-state thermal pressure $P/k_{\rm B}=nT$
versus gas density $n$, computed with the BS19 thermochemical
model\cite{Bialy2019} at fiducial $Z'=1$,
$\zeta=10^{-16}$\,s$^{-1}$, and $I_{\rm UV}=1$. The curve is
S-shaped: the stable warm (WNM) and cold (CNM) branches are joined
by the thermally unstable branch (UNM; dashed, $dP/dn<0$), bounded
by a pressure maximum at $\nwmax$ and a pressure minimum at
$\ncmin$. These three branches are the equilibrium solutions of
the model.
\textit{Bottom:} The same model recast as the classification
applied to $\mathcal{P}_{\rm 3D}$ across the full
$(n,I_{\rm UV})$ plane. The local density and radiation intensity
of each cell set its category: warm where only a stable WNM
solution exists ($n<\nwmax$), cold where only a stable CNM
solution exists ($n>\ncmin$), and unstable in between. The black
curves are the separators $\nwmax(I_{\rm UV})$ and
$\ncmin(I_{\rm UV})$; the top panel corresponds to the slice at
$I_{\rm UV}=1$, and the vertical guide lines link its turning
points to the corresponding points on the separators. The cold
and warm categories coincide with the model's stable branches,
and we refer to them as thermal phases. The unstable category is
broader than the UNM branch above: it collects every cell in the
intermediate-density interval, whether on the unstable equilibrium
branch or displaced from equilibrium by dynamical processes, the
latter not lying on the equilibrium curve at all. We therefore
refer to the intermediate gas as a category rather than a phase.
Off-equilibrium gas outside this interval is not separately
identified by the classification.}
\label{fig:phase_classification}
\end{figure}
\clearpage
\begin{figure}[H]
\centering
\includegraphics[width=0.85\textwidth]{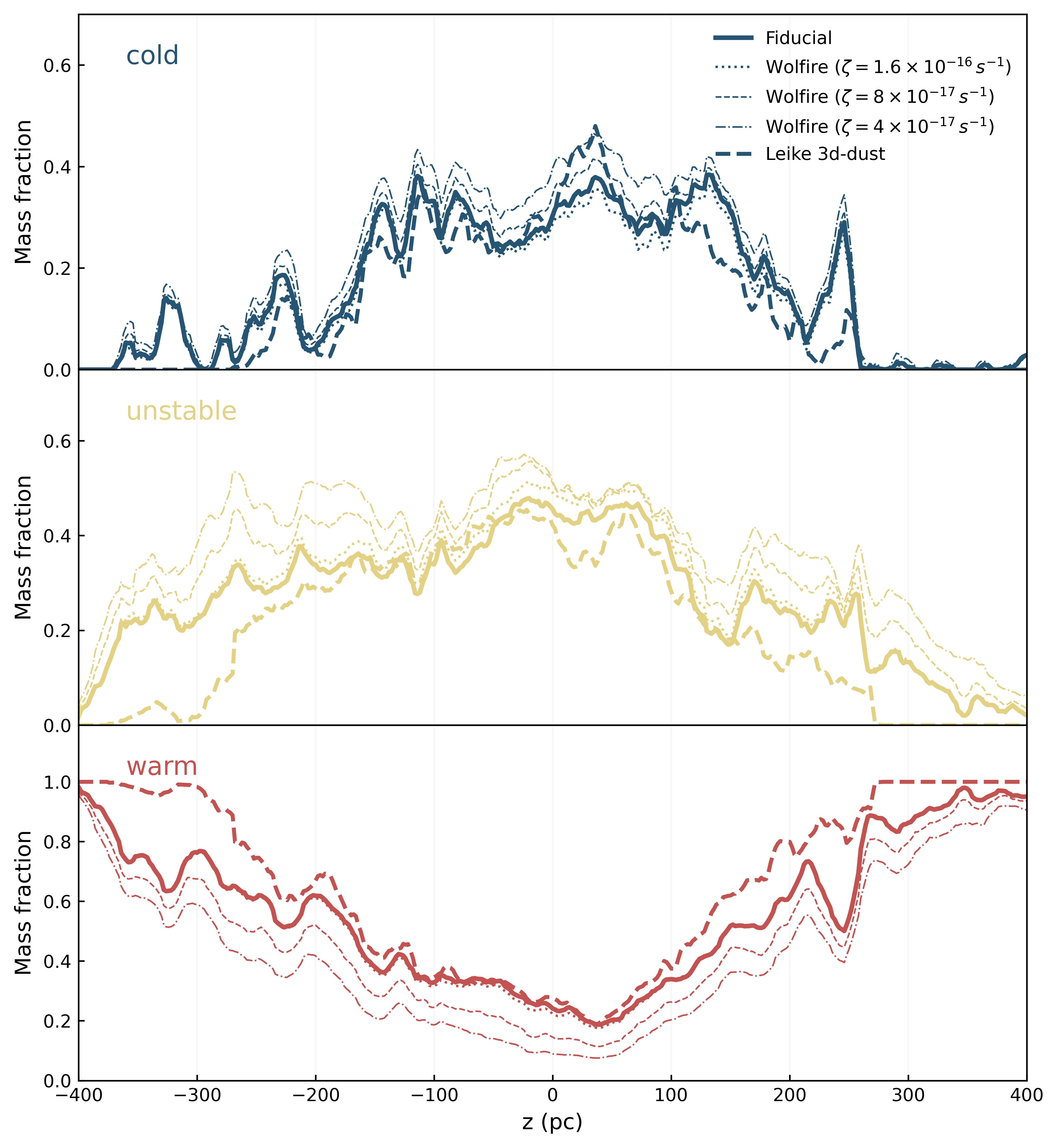}
\caption{\textbf{Robustness of phase fractions to model and dust
map choices.}
Vertical profiles of the cold (top), unstable (middle), and warm
(bottom) mass fractions for different thermochemical models and
dust maps. Thick solid: fiducial BS19 model applied to the E24
dust map. Dotted, dashed, and dot-dashed: Wolfire-type models
applied to the E24 dust map at
$\zeta=1.6 \times 10^{-16}$, $8 \times 10^{-17}$, and
$4 \times 10^{-17}$\,s$^{-1}$, respectively. Thick dashed: BS19
model applied to the Leike et al.\ dust map with the FUV field
recomputed self-consistently. All configurations reproduce the
same qualitative behaviour; quantitative differences in the
central $|z| \leq 150$\,pc layer are
$\delta f_C \approx 0.04$, $\delta f_U \approx 0.05$, and
$\delta f_W \approx 0.07$.}
\label{fig:alt_models}
\end{figure}

\begin{figure}[H]
\centering
\includegraphics[width=0.6\textwidth]{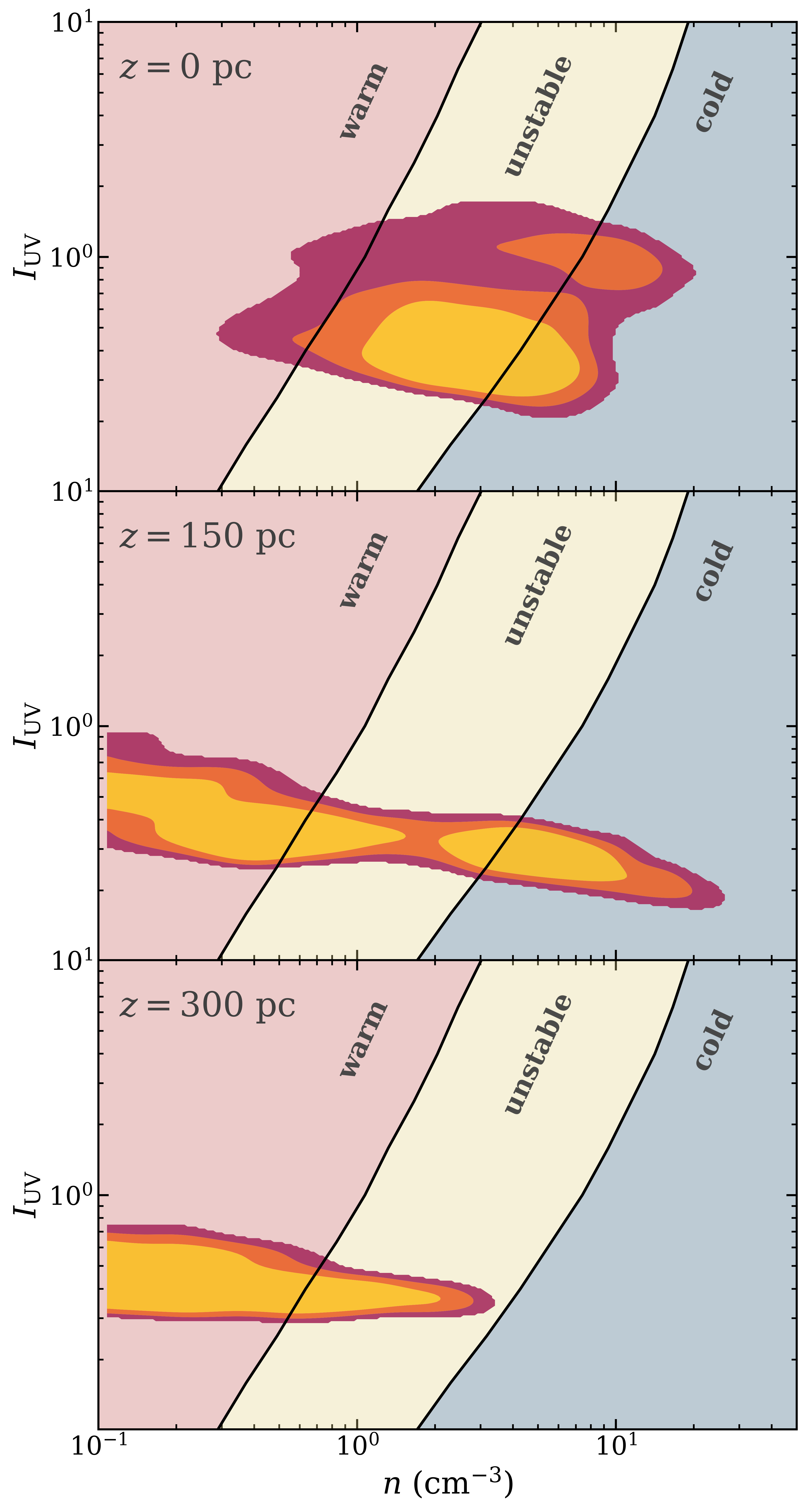}
\caption{\textbf{Vertical evolution of the joint $(I_{\rm UV},n)$
distribution.}
Mass-weighted 2D PDF of gas in the $(I_{\rm UV},n)$ plane at
three heights: $z=0$, 150, and 300\,pc. Contours enclose regions
containing the bulk of the mass in each slice. The curves show
the $\nwmax(I_{\rm UV})$ and $\ncmin(I_{\rm UV})$ phase
separators from the thermochemical model, defining the cold,
unstable, and warm regions. The distribution is extended
along the $\log n$ axis but compact in $I_{\rm UV}$, indicating
that density fluctuations dominate the phase structure while the
FUV field remains comparatively narrow. With increasing $|z|$,
the distribution shifts toward lower densities, progressively
vacating the cold region in favour of the unstable and warm
categories, consistent with Fig.~\ref{fig:phases_z}. A weak
positive correlation between $n$ and $I_{\rm UV}$ at each height
reflects the spatial proximity of dense gas to OB associations
that locally enhance the FUV field; FUV shielding by dense gas
would produce an anti-correlation but is sub-dominant in our
volume given the narrow dynamic range of $I_{\rm UV}$ relative to
$n$. A movie of the evolution with $z$ is available online at
\href{https://sb2580.github.io/phases_IUV_n_video/}{https://sb2580.github.io/phases\_IUV\_n\_video/}.}
\label{fig:iuv_n_pdf}
\end{figure}

\clearpage
\begin{table}[H]
\centering
\caption{Summary statistics of gas density, UV radiation intensity, and thermal phase mass fractions at different heights above and below the disk midplane.}
\label{tab:statistics_and_phases}
\small
\begin{tabular}{llccccccccc}
\toprule
Quantity & Statistic & $z{=}{-}350$ & $z{=}{-}250$ & $z{=}{-}150$ & $z{=}{-}50$ & $z{=}0$ & $z{=}{+}50$ & $z{=}{+}150$ & $z{=}{+}250$ & $z{=}{+}350$ \\
& & pc & pc & pc & pc & pc & pc & pc & pc & pc \\
\midrule
\multirow{5}{*}{$n$ (cm$^{-3}$)} & Mean & 0.054 & 0.13 & 0.28 & 0.54 & 0.71 & 0.64 & 0.12 & 0.083 & 0.043 \\
 & Median & 0.026 & 0.045 & 0.072 & 0.16 & 0.17 & 0.11 & 0.034 & 0.029 & 0.026 \\
 & Std & 0.17 & 0.39 & 1.1 & 1.6 & 1.7 & 1.8 & 0.55 & 0.32 & 0.069 \\
 & P$_{15}$ & 0.012 & 0.015 & 0.022 & 0.024 & 0.016 & 0.011 & 0.01 & 0.011 & 0.012 \\
 & P$_{85}$ & 0.065 & 0.16 & 0.33 & 0.82 & 1.3 & 1.1 & 0.15 & 0.091 & 0.062 \\
[2ex]
\multirow{5}{*}{$\log n$ (cm$^{-3}$)} & Mean & -1.5 & -1.3 & -1.1 & -0.83 & -0.83 & -0.95 & -1.4 & -1.5 & -1.6 \\
 & Median & -1.6 & -1.3 & -1.1 & -0.79 & -0.78 & -0.97 & -1.5 & -1.5 & -1.6 \\
 & Std & 0.4 & 0.53 & 0.61 & 0.73 & 0.86 & 0.87 & 0.57 & 0.5 & 0.37 \\
 & P$_{15}$ & -1.9 & -1.8 & -1.7 & -1.6 & -1.8 & -1.9 & -2 & -2 & -1.9 \\
 & P$_{85}$ & -1.2 & -0.79 & -0.49 & -0.085 & 0.11 & 0.035 & -0.83 & -1 & -1.2 \\
[2ex]
\multirow{5}{*}{$I_{\rm UV}$} & Mean & 0.49 & 0.54 & 0.65 & 0.83 & 0.99 & 0.77 & 0.59 & 0.52 & 0.47 \\
 & Median & 0.49 & 0.53 & 0.61 & 0.72 & 0.79 & 0.67 & 0.58 & 0.51 & 0.47 \\
 & Std & 0.058 & 0.095 & 0.54 & 0.92 & 2.3 & 1 & 0.57 & 0.27 & 0.086 \\
 & P$_{15}$ & 0.44 & 0.45 & 0.48 & 0.47 & 0.46 & 0.42 & 0.42 & 0.4 & 0.38 \\
 & P$_{85}$ & 0.55 & 0.63 & 0.81 & 1.2 & 1.4 & 0.99 & 0.73 & 0.63 & 0.56 \\
[2ex]
\multirow{5}{*}{$\log I_{\rm UV}$} & Mean & -0.31 & -0.27 & -0.21 & -0.13 & -0.093 & -0.17 & -0.25 & -0.3 & -0.33 \\
 & Median & -0.31 & -0.28 & -0.22 & -0.14 & -0.1 & -0.17 & -0.24 & -0.29 & -0.33 \\
 & Std & 0.05 & 0.071 & 0.11 & 0.19 & 0.24 & 0.2 & 0.13 & 0.092 & 0.078 \\
 & P$_{15}$ & -0.36 & -0.35 & -0.32 & -0.33 & -0.34 & -0.38 & -0.37 & -0.4 & -0.42 \\
 & P$_{85}$ & -0.26 & -0.2 & -0.093 & 0.061 & 0.14 & -0.006 & -0.14 & -0.2 & -0.25 \\
\midrule
\multicolumn{11}{c}{\textbf{Thermal Phase Mass Fractions}} \\
\midrule
$f_C$ &  & 0.023 & 0.07 & 0.26 & 0.25 & 0.28 & 0.32 & 0.21 & 0.11 & 0 \\
$f_U$ &  & 0.19 & 0.34 & 0.34 & 0.44 & 0.51 & 0.49 & 0.22 & 0.23 & 0.048 \\
$f_W$ &  & 0.79 & 0.59 & 0.4 & 0.31 & 0.21 & 0.19 & 0.56 & 0.66 & 0.95 \\
\midrule
\multicolumn{11}{c}{\textbf{Mean fractions for $|z| \leq 150$\,pc:}\quad $\langle f_C \rangle = 0.27$\quad $\langle f_U \rangle = 0.41$\quad $\langle f_W \rangle = 0.32$} \\
\bottomrule
\end{tabular}
\vspace{6pt}
\tablecomments{ \\
Upper section: mean, median, standard deviation (Std), and 15th and 85th percentiles (P$_{15}$, P$_{85}$) for gas density $n$, logarithmic density $\log n$, UV radiation intensity $I_{\rm UV}$, and logarithmic UV intensity $\log I_{\rm UV}$.\\
Lower section: Mass fractions of the three thermal categories at selected heights and averaged over $|z| \leq 150$\,pc. Mass fractions sum to unity at each height.
}
\end{table}
\end{document}